\documentclass[12pt,preprint]{aastex}
\usepackage{graphics}
\usepackage{aastexug}  
\usepackage{natbib}

\newcommand{\ocen}{$\omega$~Cen}
\def\Msun{$\rm M_\odot$}
\def\Mc{$\rm M_{\rm c}$}

\def\Teff{$T_{\rm eff}$}

\def\simgt{\lower.5ex\hbox{$\; \buildrel > \over \sim \;$}}
\def\simlt{\lower.5ex\hbox{$\; \buildrel < \over \sim \;$}}
\def\ltsima{$\; \buildrel < \over \sim \;$}
\def\gtsima{$\; \buildrel > \over \sim \;$} 
\def\lsim{\lower.5ex\hbox{\ltsima}} 
\def\gsim{\lower.5ex\hbox{\gtsima}} 
\def\msun{${\rm M_\odot}$} 
 
\begin{document} 
 
\title{Terzan 5: an alternative interpretation for the split horizontal branch}
 
\author{F. D'Antona\altaffilmark{1}, P. Ventura\altaffilmark{1}, V. Caloi\altaffilmark{2}, 
 A. D'Ercole\altaffilmark{3}, E. Vesperini\altaffilmark{4}, R. Carini\altaffilmark{1,5}, 
M. Di Criscienzo\altaffilmark{1} 
}
\affil{
\altaffilmark{1}
INAF--Osservatorio Astronomico  di Roma, via Frascati
33, I-00040 Monteporzio, Italy; 
\\
\altaffilmark{2}
INAF--IASF--Roma, via Fosso del Cavaliere 100, I-00133 Roma, Italy; 
\\
\altaffilmark{3}
INAF--Osservatorio Astronomico  di Bologna, via Ranzani 1, I-40127 BOLOGNA (Italy)\\
\altaffilmark{4}
Department of Physics, Drexel University, Philadelphia, PA 19104, USA\\
\altaffilmark{5}
Department of Physics, Universit\`a di Roma - La Sapienza, Roma, Italy; 
\\
}

\begin{abstract}  
We consider the horizontal branch (HB) of the Globular Cluster Terzan~5, recently shown to be split 
into two parts, the fainter one ($\delta M_K \sim 0.3$mag) 
having a lower metallicity than the more luminous. Both features 
show that it contains at least two stellar populations. 
The separation in magnitude has been ascribed to an age difference of 
$\sim$6~Gyr and interpreted as the result of an atypical evolutionary history
for this cluster. We show that the observed HB
morphology is also consistent with a model in which the bright
HB  is composed of second generation stars that are metal enriched and
with a helium mass fraction larger (by $\delta Y \sim$0.07) than
that of first generation stars populating the fainter part of the
HB. Terzan 5 would therefore be anomalous, compared to most ``normal"
clusters hosting multiple populations, only because its second
generation is strongly contaminated by supernova ejecta; the previously
proposed prolonged period of star formation, however, is not
required. 
The iron enrichment of the bright HB can be ascribed either to 
contamination from Type Ia supernova ejecta of the low--iron, helium rich, ejecta of the massive
asympotic giant branch stars of the cluster, or to its mixing 
with gas, accreting on the cluster from the environment, that has been
subject to fast metal enrichment due to its proximity with the galactic bulge.
The model here proposed requires only a small age difference, of $\sim$100~Myr.
\end{abstract}


\section{Introduction}
\label{sec:intro}
Recent years have witnessed exciting developments both in the observations and 
theoretical modelling of the abundance star to star variations within most of 
the well studied Globular Clusters (GCs). In most GCs star to star abundance variations are limited
to the light elements that are susceptible to abundance changes from 
proton-capture reactions, such as the pp, CN, ON, NeNa, and MgAl cycles \citep[for the most
recent spectroscopic survey, see, e.g.][]{carretta2009a,carretta2009b}, but a few clusters, such
as M22 \citep{marino2009}, or perhaps NGC 1851 \citep{han2009} are now known to exhibit variations 
in heavier elements \citep[see also][]{carretta2009ferro}, and, 
more than the others, in $\omega$Cen the heavy elements spreads 
\citep[e.g., among others][]{norrisdacosta1995},
and the HR diagram morphologies clearly show that we are dealing with several 
stellar generations, enriched by the supernova (SN) ejecta \citep[e.g.][]{sollima2005, 
villanova2007}. In addition, the cluster M~54, 
immersed in the nucleus of the Sagittarius dwarf galaxy, 
presently disrupting in our Galaxy \citep{ibata1994,bellazzini2008} has been 
recently found to show a metallicity spread similar to \ocen\ \citep{carrettam542010}

Concerning the spread in light elements, their observation 
at the turnoff and among the subgiant stars 
\citep[e.g.,][]{gratton2001} showed that these anomalies must be attributed 
to self--enrichment occurring at the first stages of the 
life of the cluster. The most peculiar finding of the latest years is the presence of
a very helium rich population in the most massive clusters: 
this is revealed by the presence of multiple main
sequences in \ocen\ and NGC~2808, indicating a helium content
Y=0.38--0.40 \citep{norris2004,dantona2005,piotto2007}, and by the extreme 
morphology of the horizontal branch (HB) of the two massive clusters  NGC~6388 and
NGC~6441. In these latter clusters, a red clump is expected as HB, due to their
large metallicity \citep[{[Fe/H]$\sim$--0.4}, see][]{carretta2009ferro}. On the contrary, 
their HB is extended towards the blue,
and the RR Lyr variables have so long periods that they must be highly overluminous. \cite{caloi2007}
and \cite{dc2008} show that the HB morphology and RR Lyr's periods of these two clusters
may be explained if a large fraction of the HB stars have Y$\simgt$0.35.
The presence of much more moderate helium spreads is probable in most of the other smaller clusters
\cite[e.g.][]{dc2008}. 
The quasi--constancy of heavy metals in most GCs leads to hypothesize that 
the abundance variations must be due to very peculiar chemical evolution, not or scarcely affected by
SN ejecta, but involving formation of a ``second generation" (SG) of stars from matter processed
into the ``first generation" (FG) stars. On the other hand, the numerical consistency of the SG is so 
high \citep[$>$50\%][]{dc2008,carretta2009a} that any formation model must include the hypothesis that
the mass contained in FG stars ---that contribute to this second phase of star formation--- 
is {\sl initially} much larger than the mass present today in the cluster. 
Models for the formation of these multiple generations are 
still in their infancy. We can divide them into two main categories: the models in which clusters
are born inside dwarf galaxies, so that the polluting matter on the forming GC comes from a much
larger environment \citep[e.g.][]{bekki2007}, and the models in which there
is an initial cluster 10--20 times more massive than todays'. In the latter case, 
the SG forms from the ejecta of the 
FG stars mainly in the central cluster parts, and the first dynamical phases of evolution 
lead to a preferential loss of the FG stars \citep{dercole2008}. In these models, it is very 
difficult to accomodate large age differences between the first and second generation stars.

Consequently, the recent observations of the color magnitude diagram features and chemistry of 
Terzan 5 may constitute a benchmark in our understanding of GC formation.
In fact, \cite{ferraro2009} show that the cluster HB stars are divided into two
clumps separated by $\delta$M$_K \sim 0.3$mag, and that the more luminous stars have a much larger
iron content ([Fe/H]$\sim$+0.3$\pm$0.1) with respect to the lower HB ([Fe/H]$\sim$--0.2$\pm$0.1). 
This result shows that the evolution of this cluster is atypical, and that matter forming the SG stars 
(populating the upper HB clump), has been affected by SN contamination, 
as it occurred in \ocen. 
On the other hand, \cite{ferraro2009}, comparing the HB data to stellar isochrones of the correct metallicity,
conclude that the SG must be $\sim$6~Gyr younger than the FG. This huge age difference is very
difficult to be understood in any formation framework, and this would be the first evidence for such a young 
age among bulge stars and clusters \citep[e.g.][]{feltzing2000, origlia2008}.

We re--examine the problem and show that
the HB morphology can also be explained by two coeval populations having an helium difference of 
$\delta$Y$\sim 0.07$, thus reaching a value not as extreme as in the cases quoted above,
so its formation does not present particular problems (Sect. \ref{sec:disc}).
At the super--solar metallicity
of the SG of Terzan~5 the possible helium enrichment in the SG does not produce
a blue extension of the HB. In Sect. \ref{sec:disc} we discuss some possibilities for the 
chemical evolution of the cluster.
We finally remark that the different space distribution of the two populations might
imply a mass difference between them, and that further
observations and dynamical modelling may allow to choose between models based on age-- 
or helium--differences.

\section{The stellar models}
We computed evolutionary tracks, isochrones and HB tracks 
for a metallicity Z=0.01 (having [Fe/H]$\sim$--0.2 for a solar--scaled mixture) 
and helium content Y=0.26, representing the FG of Terzan 5 and its lower HB clump, 
and models and isochrones for Z=0.03 ([Fe/H]$\sim$+0.3), Y=0.29
and Y=0.40. The standard inputs of our evolutionary code ATON are used \citep{ventura2009a}. 
We adopt the opacities by \cite{ferguson2005} 
at temperatures lower than 10000 K and the OPAL opacities in the version 
documented by \cite{iglesias1996}. 
The mixture adopted is solar--scaled and follows the element distribution by \cite{GS98}. 
For comparison, we also computed models for iron content [Fe/H]$\sim$--0.2\ and [$\alpha$/Fe]=0.4. 
Electron conduction opacities follow the treatment by \cite{potekhin1999}, and
are harmonically added to the radiative opacities.
The HB models are constructed by assuming the core helium mass derived from the red giant evolution of
masses evolving in the range of ages 10--13Gyr for the given Z and Y. 
The values assumed are then \Mc=0.4793\msun\ for Z=0.01 and Y=0.26; \Mc=0.464\msun\ for Z=0.03, Y=0.29
and \Mc=0.448\msun for Z=0.03 and Y=0.40.
From the isochrones we derive the mass at the tip of the giant branch for each age and chemical composition.
These values are used to build up synthetic models of HB, following the procedure described by \cite{dc2008}. 
The mass lost along the red giant branch and its dispersion are fixed for each given age, in order to
derive the distribution of HB masses. A further random extraction of the age within the HB lifetime
allows to populate the synthetic HB. We fix the FG population at the chemistry Y=0.26, Z=0.01, and
the SG at Z=0.03. The helium content of the SG was allowed to vary between 0.29 and 0.40, but
we assume the same mass loss for both populations.
The theoretical values of luminosity and \Teff\ are converted into the Johnson--Bessell 
system by means of \cite{bcp1998}. 

\section{The HB morpology of Terzan~5: two possible interpretations}
We start by discussing in this section an example that just serves to
illustrate how an increase in the helium abundance can explain the difference in K magnitude
between the two clumps in Terzan 5.
Figure 1 shows a representation of the results that allows to graphically visualize
both \cite{ferraro2009} conclusion and a different interpretation of the data proposed here.
In the right panel of Fig.1 we shows the HB mass versus age relations 
\footnote{We plot the mass at the tip of the red giant branch, 
that is a decreasing function of age, minus 0.3\msun, in order to provide an example 
of what may be the masses populating the HB.} and on the left panel
we plot the HB mass versus the absolute magnitude M$_K$\ of the ZAHB.
Let us try to understand why we need $\sim 6$Gyr of age difference, if we
assume (Z=0.01, Y=0.26) as composition of the faint clump and (Z=0.03, Y=0.29)
for the brighter one \citep{ferraro2009}. Assume that the age of the faint clump is 12.5Gyr. The
evolving mass on the HB is represented by the point labelled A 
(age=12.5Gyr and M=0.646\msun). This mass on the ZAHB has M$_K$=--1.16 (point B). 
Now we know that the other clump is $\sim$0.3mag brighter, so that
we shift to M$_K$=--1.46 (point C). This magnitude needs a HB mass of 0.9\msun\ (point D), if
we are dealing with models of Y=0.29, Z=0.03. The {\it age} corresponding to this evolving mass
\footnote{This is the age, if the mass loss does not increase with metallicity, otherwhise it must
be considered an upper limit to the age} is then 7Gyr (point E). Thus the brighter clump is
at least 5.5Gyr younger than the fainter clump, as \cite{ferraro2009} find.
\begin{figure}[t]
\begin{center}
\includegraphics[width=8cm]{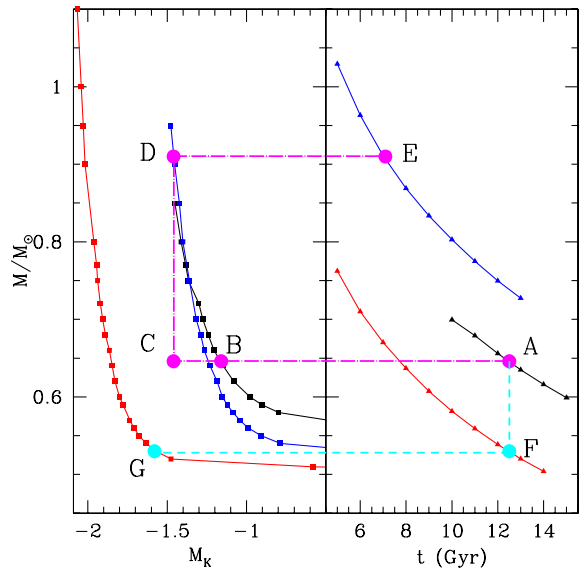}
\caption{Right panel: the curves with triangles represent the mass versus age relations for the three 
different compositions (Y=0.29, Z=0.03: top) (Y=0.26, Z=0.01: middle) (Y=0.40, Z=0.03 bottom). The mass
is the red giant mass evolving at each age, minus 0.3\msun, that represents an average mass lost
on the red giant branch. Thus the relations represent the average HB mass as function of age.
Left panel: the curves with squares represent the K--magnitude as a function of the mass in the ZAHBs of 
(Y=0.26, Z=0.01: top); (Y=0.29, Z=0.03, middle); (Y=0.40, Z=0.03: bottom). The letters label
how we can have two populations differing by $\delta$M$_K \sim 0.3$mag by increasing Y and Z
(from A to G, dashed path) or by increasing Z and decreasing the age (from A to E, dash--dotted path) 
as explained in the text. }
\label{f1}
\end{center}
\end{figure}
Let us suppose now that the two populations are practically coeval, and that the brighter
clump corresponds to a population having Z=0.03 and Y=0.40. In this case, we must have an evolving
mass of 0.528\msun\ (point F) that has M$_K$=--1.58 on the ZAHB (point G). 
In this case, the gap in M$_K$\ between the two populations is the magnitude 
difference between points B and G ($-1.16+1.58=$0.42mag).
This example shows that a helium increase can easily explain the magnitude difference between the two
clumps, without requiring an age difference.
\begin{figure}[t]
\begin{center}
\includegraphics[width=8cm]{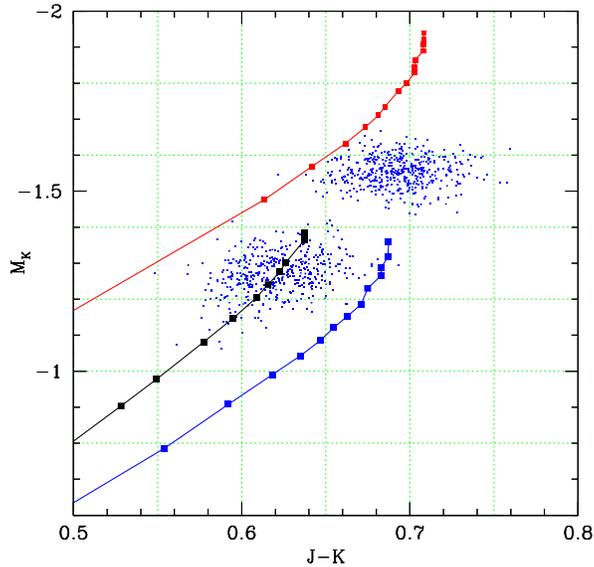}
\caption{The dots represent the simulation of the HB in Terzan 5 by assuming a population having
Y=0.26 and Z=0.01 (fainter clump) and a brighter HB with Y=0.33 and Z=0.03 (upper clump). 
The three lines with squares represent the ZAHBs of our models: Z=0.01 at bottom left,
Z=0.03, Y=0.29 at bottom right, and Z=0.03, Y=0.40 (upper line).}
\label{f2}
\end{center}
\end{figure}  

Among the many simulations of the HB, produced to understand the role of the different parameters, 
Figure \ref{f2} shows the simplest one that reproduces the gap of M$_K \sim 0.3$mag between the
two populations. We assume Y=0.33 for the bright clump. All stars have 
age 11Gyr, mass loss along the red giant branch $\delta$M=0.28\Msun, 
with dispersion $\sigma$=0.025\msun.
Both the color and luminosity difference of the clumps are reasonably reproduced.
The choice of models, age and helium enhancement is not unique. The luminous clump can also be composed
of stars with Y varying in the range Y=0.32--0.34, born from ejecta with different helium.
If $\delta$Y$<$0.07 between the two populations, however, the M$_K$\ gap can not be reproduced.
We have assumed for both the FG and SG a solar scaled composition, a reasonable choice for the SG,
if its higher metallicity is due to Type Ia SN contamination, that pollute
mostly with iron and bring the composition towards the solar--scaled abundances. If the FG is 
instead $\alpha$--enhanced, and we assume for it [Fe/H]=$-0.2$\ and [$\alpha$/Fe]=0.4, 
our models show that the ZAHB shifts to redder J--K, by $\sim$0.03mag, and
the color difference of the two clumps is reduced. A small reduction of the iron content,
within the range allowed by the measurements errors, would reproduce again the color fit.

The iron content of the upper clump may be not unique for all stars, if it is a 
result of non--homogeneous contamination of the matter forming the SG stars. In this case, a proper
interpolation between HBs of different [Fe/H] must be taken into account, but a similar result
will be obtained.

Notice that both the FG and the SG clumps ``stay in the red", 
while the other high metallicity clusters, 
NGC~6388 ([Fe/H]$\sim$--0.40) and NGC~6441 ([Fe/H]$\sim$--0.33) \citep{carretta2009ferro},
have HBs extended towards large \Teff\ thanks to the much larger Y of their SG.
In fact, the iron content of Terzan~5 is {\it much larger}, and 
the H--burning shell maintains a giant structure and a large radius for HB stars even of
relatively small mass. Our models show that we need 0.53\Msun\ to depart towards 
larger \Teff\ for Y=0.29 and 
0.51\Msun\ for Y=0.40. Notice that a HB mass of 0.56\Msun\ occupies the RR~Lyr pulsation strip for the 
FG chemistry. A single RR~Lyr and a few blue HB stars have been found in the cluster \citep{cohn2002}: 
these may well represent the tail of the mass distribution of HB stars populating 
either of the red clumps, and it would be interesting to know which one.

\section{The chemical evolution of Terzan 5: type II or type Ia SN enrichment?}
\label{sec:disc}
We have shown that the HB of Terzan 5 may be explained either by two
populations with an age difference of ~6 Gyr or with two populations
that are approximately coeval ($\delta$(age)$\sim$100~Myr) but have a different
helium content ($\delta$Y$\sim$0.07). The latter interpretation would put
the formation of Terzan~5 within the  theoretical framework suggested for
the formation of most GCs with multiple populations \citep[e.g.][]{dercole2008}.
However, while in most clusters SG stars have iron abundance similar to FG stars, the
iron enhancement in the second generation population of Terzan 5
introduces a new ingredient in that scenario and requires 
the identification of the source of metal rich gas.

We tentatively identify the helium rich gas with the massive--AGB and super--AGB ejecta 
\citep{ventura2002,pumo2008}, and compute the evolution of these stars for the FG composition (Z=0.01), 
following \cite{vd2009}. The smallest star igniting carbon in conditions of
semi-degeneracy (super--AGB evolution) is the 7.5\Msun, evolving at 50Myr. The AGB ejecta have helium
abundance between Y=0.36 (7\msun) and Y=0.32\Msun\ (5\msun), so masses down to $\sim$5\msun\ 
(evolving at $\sim$100Myr) can fit the SG requirements. We have $\sim$100Myr to pollute the
AGB ejecta with iron rich material. 
A change in the iron content from [Fe/H]$\sim$--0.2$\pm$0.1 to [Fe/H]$\sim$+0.3$\pm$0.1, 
assuming Z$_\odot$=0.018, an iron mass fraction f=0.074 from \cite{GS98}, and solar--scaled 
compositions, means an increase in the iron mass fraction by 1.8$\times 10^{-3}$ in all the SG mass. 
Assuming that this mass is half of the total mass of the cluster today, say 2.5$\times 10^5$\Msun, 
the requirement is 456\Msun\ of iron.
We can think of two ways of forming an SG both helium and metal enriched. 

1) The helium enriched ejecta are directly polluted by iron
produced by supernovae belonging to the cluster itself.  
If the source of the helium enrichment are the massive stars, as in \cite{decressin2007} model,
the source of the metal enrichment must be the SNII. Each SNII produces only $\sim$0.07\Msun\ of iron
\citep{hamuy2003}, so we would need 6500 SNII explosions to reach the required iron content, an 
occurrence that would very likely destroy the cluster, and will not be considered any longer.
If the source of helium enrichment are the massive AGBs, like in the dynamical model 
proposed by \cite{dercole2008}, for Terzan~5 we must further
require that the second stage of star formation is not halted by the injection of energy from the
SN explosions, that generally is regarded as the cause of the end of 
the SG formation epoch. Below 7.5\Msun, as remarked above, stars evolve into carbon oxygen
white dwarfs, so that, some time after, SNIa can begin exploding in the cluster. 
Each SNIa injects in the intracluster medium $\sim$0.6\Msun\ of iron, so we need 760 SNIa explosions. 
A rate of 1 SNIa every 50000yr can provide this iron in 38Myr, when the AGB ejecta are still as helium 
rich as needed. If the SG will result to be homogeneous in iron, the model requires that
the helium rich AGB gas accumulates, and that the SNIa
explode polluting the AGB gas, but not destroying its accumulation.
The above back of the envelope computation shows that 
the requirements for the increase of the iron content may be {\it in principle} fulfilled, although
with many caveats. In this scenario we expect that in the SG [$\alpha$/Fe]$\sim$0.
As for the other elemental abundances, this model can be easily falsified, as it predicts that
the SG {\it should not} show the enhanced [Na/Fe] signature typical of GCs. In fact, the AGB
gas may well be sodium enhanced with respect to its iron content [Fe/H]$\sim -0.2$, but the iron rich
mixture forming the SG will necessarily have a low [Na/Fe].

2) The SG matter consists of the low--iron gas,
ejected from the cluster AGBs, mixed with the {\it environment} gas accreting on the cluster from 
the neighborhood regions. Due the proximity to the galactic bulge, this gas has
been subject to fast metal enrichment, and its iron may be largely supersolar
even at a very early epoch, depending on the scenario of formation and evolution of the
bulge \citep[e.g.][]{wysegilmore1992}. Mixing of the hot--CNO processed gas forming any GC SG 
with ``pristine" gas is a common requirement of models in all clusters showing
the signature of the sodium -- oxygen anticorrelation \citep[e.g.][]{prantzos2007, bekki2007}. 
If the AGB gas is diluted with a similar quantity of accreted metal--rich gas, we require that 
[Fe/H]$\sim$0.44 in the accreting gas. Such a metal rich matter itself will have a  
larger than solar helium content, say Y$\simgt$0.30 
\citep[Matteucci, private communication, see also][]{renzini1999}, so that the
very helium rich composition of the mixture forming the SG is a tenable hypothesis too.
On the contrary, it is difficult to make predictions on abundances of the other elements 
since they will come out from the above described mixture of hot--CNO--processed matter (e.g. with
low [O/Fe] and large [Na/Fe]) with the bulge iron rich gas. The specific abundances in the
bulge gas depend on its precise evolutionary history, still not fully understood \citep{matteucci1999,
lecureur2007,ballero2007}. 

\section{Conclusions}
We have shown that the split HB of Terzan~5 can be interpreted as due to two populations differing 
in helium content and metallicity, and not much different in age ($\delta$(age)$\simlt$100Myr). 
Massive AGB stellar models
for the chemistry of the FG are compatible with the required helium enhancement, but we need 
that 1) either the AGB matter itself is strongly polluted by SNIa ejecta, 
before the second stage of star formation begins, or 2) the AGB matter is diluted with
accreted gas, fastly processed to very high metallicity in the bulge stellar environment. 
This suggestion may help to understand the ``true" birth of the double population 
of this cluster, maybe as a mix of age and helium difference in the subsequent star formation events.

We conclude by pointing out that the two alternative scenarios (age or helium difference)
predict different values for the bright HB and faint HB masses. Specifically while
in the merging scenario the younger age of the bright HB implies that this
population would be $\sim$0.25\msun\ more
massive than the faint HB (in Fig.\ref{f1}, the mass difference between
points C and D), in the scenario proposed in this paper the
two populations would be almost coeval and their red giant progenitors would have only a small
mass difference (in the example of Fig. \ref{f2}, 
M=0.996\msun\ for the bright HB and  M=0.979\msun for the faint HB). 
Further dynamical modelling will help to shed further light on the
plausibility of the two scenarios and on the possible dynamical
histories leading to the observed differences in the spatial
distribution of the two populations.
 
\section{Acknowledgments} 
This work has been supported through PRIN MIUR 2007 
``Multiple stellar populations in globular clusters: census, characterization and
origin". We thank F. Ferraro for useful information on the Terzan~5 observations and
F. Matteucci for information on the possible helium content of the bulge.


\begin{thebibliography}{}

\bibitem[Ballero et al.(2007)]{ballero2007} Ballero, S.~K., Matteucci, F., 
Origlia, L., \& Rich, R.~M.\ 2007, \aap, 467, 123 

\bibitem[Bekki et al.(2007)]{bekki2007} Bekki, K., Campbell, 
S.~W., Lattanzio, J.~C., \& Norris, J.~E.\ 2007, \mnras, 377, 335 

\bibitem[Bellazzini et al.(2008)]{bellazzini2008} Bellazzini, M., et 
al.\ 2008, \aj, 136, 1147 

\bibitem[Bessell et al.(1998)]{bcp1998} Bessell, M.~S., Castelli, F., \& Plez, B.\ 
1998, \aap, 333, 231 

\bibitem[Caloi \& D'Antona(2007)]{caloi2007} Caloi, V., \& D'Antona, F.\ 2007, \aap, 463, 949 

\bibitem[Carretta et al.(2009a)]{carretta2009a} Carretta, E., et al.\ 2009a, \aap, 505, 117 

\bibitem[Carretta et al.(2009b)]{carretta2009b} Carretta, E., Bragaglia, A., Gratton, R., 
\& Lucatello, S.\ 2009b, \aap, 505, 139 

\bibitem[Carretta et al.(2009c)]{carretta2009ferro} 
Carretta, E., Bragaglia, A., Gratton, R., D'Orazi, V., \& Lucatello, S.\ 2009c, \aap, 508, 695 

\bibitem[Carretta et al.(2010)]{carrettam542010} Carretta, E., et al.\ 
2010, arXiv:1002.1963, ApJL, in press 

\bibitem[Cohn et al.(2002)]{cohn2002} Cohn, H.~N., Lugger, 
P.~M., Grindlay, J.~E., \& Edmonds, P.~D.\ 2002, \apj, 571, 818 

\bibitem[D'Antona \& Caloi(2004)]{dantona2004} D'Antona, F.~\& Caloi, V.\ 2004, 
ApJ, 611, 871 

\bibitem[D'Antona et al.(2005)]{dantona2005} D'Antona, F., Bellazzini, M., 
Caloi, V., Pecci, F.~F., Galleti, S., \& Rood, R.~T.\ 2005, \apj, 631, 868 

\bibitem[D'Antona \& Ventura(2007)]{dv2007} D'Antona, F., \& Ventura, P.\ 2007, \mnras, 379, 1431 

\bibitem[D'Antona \& Caloi(2008)]{dc2008} D'Antona, F., \& Caloi, V.\ 2008, 
MNRAS, 390, 693

\bibitem[Decressin et al.(2007)]{decressin2007} Decressin, T.,  
Meynet, G., Charbonnel, C., Prantzos, N., \& Ekstr{\"o}m, S.\ 2007, A\&A, 
464, 1029 

\bibitem[D'Ercole et al.(2008)]{dercole2008} D'Ercole, A., 
Vesperini, E., D'Antona, F., McMillan, S.~L.~W., \& Recchi, S.\ 2008, \mnras, 391, 825 

\bibitem[Feltzing \& Gilmore(2000)]{feltzing2000} Feltzing, S., \& Gilmore, G.\ 2000, \aap, 355, 949 

\bibitem[Ferguson et al.(2005)]{ferguson2005} Ferguson  J. W., Alexander  D. R., Allard  F.  et al., 2005, ApJ, 623, 585

\bibitem[Ferraro et al.(2009)]{ferraro2009} Ferraro, F.~R., et al.\ 
2009, \nat, 462, 483 

\bibitem[Gratton et al.(2001)]{gratton2001} Gratton, R.~G.~et al.\ 2001, \aap, 
369, 87

\bibitem[Gratton, Sneden \& Carretta(2004)]{gratt-annualrev} Gratton, G., 
Sneden, C., \& Carretta, E. 2004, \araa, 42, 385

\bibitem[Grevesse \& Sauval(1998)]{GS98} Grevesse, N., \& Sauval, A.J. 1998, SSRv, 85, 161

\bibitem[Hamuy(2003)]{hamuy2003}Hamuy, M. 2003, ApJ, 582, 905

\bibitem[Han et al.(2009)]{han2009} Han, S.-I., Lee, Y.-W., 
Joo, S.-J., Sohn, S.~T., Yoon, S.-J., Kim, H.-S., 
\& Lee, J.-W.\ 2009, \apjl, 707, L190 

\bibitem[Ibata et al.(1994)]{ibata1994} Ibata, R.~A., Gilmore, 
G., \& Irwin, M.~J.\ 1994, \nat, 370, 194 

\bibitem[Iglesias  \& Rogers(1996)]{iglesias1996}Iglesias  C. A. \& Rogers  F. J., 1996, ApJ, 464, 943

\bibitem[Lecureur et al.(2007)]{lecureur2007} Lecureur, A., Hill, V., Zoccali, M., Barbuy, B., 
G{\'o}mez, A., Minniti, D., Ortolani, S., \& Renzini, A.\ 2007, \aap, 465, 799 

\bibitem[Marino et al.(2009)]{marino2009} Marino, A.~F., Milone, A.~P., Piotto, G., 
Villanova, S., Bedin, L.~R., Bellini, A., \& Renzini, A.\ 2009, \aap, 505, 1099 

\bibitem[Matteucci et al.(1999)]{matteucci1999} Matteucci, F., Romano, D., \& Molaro, P.\ 1999, \aap, 341, 458 

\bibitem[Norris(2004)]{norris2004} Norris, J.~E.\ 2004, \apjl, 
612, L25 

\bibitem[Norris \& Da Costa(1995)]{norrisdacosta1995} Norris, J.~E., \& Da Costa, G.~S.\ 1995, 
\apj, 447, 680 

\bibitem[Origlia et al.(2008)]{origlia2008} Origlia, L., Lena, S., 
Diolaiti, E., Ferraro, F.~R., Valenti, E., Fabbri, S., 
\& Beccari, G.\ 2008, \apjl, 687, L79 

\bibitem[Piotto et al.(2005)]{piotto2005} Piotto, G., et al.\ 
2005, \apj, 621, 777 

\bibitem[Piotto et al.(2007)]{piotto2007} Piotto, G., et al.\ 
2007, ApJL, 661, L53 

\bibitem[Potekhin et al.(1999)]{potekhin1999}Potekhin, A. Y.; Baiko, D. A.; Haensel, P.; Yakovlev, D. G., 1999, A\&A, 346, 345P

\bibitem[Prantzos et 
al.(2007)]{prantzos2007} Prantzos, N., Charbonnel, C., \& Iliadis, C.\ 2007, \aap, 470, 179 

\bibitem[Pumo et al.(2008)]{pumo2008} Pumo, M.~L., D'Antona, F., 
\& Ventura, P.\ 2008, \apjl, 672, L25 

\bibitem[Renzini(1999)]{renzini1999} Renzini, A.\ 1999, \apss, 267, 357 

\bibitem[Sollima et al.(2005)]{sollima2005} Sollima, A., Pancino, 
E., Ferraro, F.~R., Bellazzini, M., Straniero, O., 
\& Pasquini, L.\ 2005, \apj, 634, 332 

\bibitem[Ventura, D'Antona, \& Mazzitelli(2002)]{ventura2002} Ventura, P., 
D'Antona, F., \& Mazzitelli, I.\ 2002, \aap, 393, 215

\bibitem[Ventura \& D'Antona(2009)]{vd2009} Ventura, P., \& D'Antona, F.\ 2009, \aap, 499, 835 

\bibitem[Ventura et al.(2009)]{ventura2009a} Ventura, P., Caloi, V., 
D'Antona, F., Ferguson, J., Milone, A., 
\& Piotto, G.~P.\ 2009, \mnras, 399, 934 

\bibitem[Villanova et al.(2007)]{villanova2007} Villanova, S., et 
al.\ 2007, \apj, 663, 296 

\bibitem[Wyse \& Gilmore(1992)]{wysegilmore1992} Wyse, R.~F.~G., \& Gilmore, G.\ 1992, \aj, 104, 144 

\bibitem[Yong et al.(2003)]{yong2003} Yong, D., Grundahl, F., Lambert, D.~L., 
Nissen, P.~E., \& Shetrone, M.~D.\ 2003, \aap, 402, 985 

\bibitem[Yong et al.(2005)]{yong2005} Yong, D., Grundahl, F., 
Nissen, P.~E., Jensen, H.~R., \& Lambert, D.~L.\ 2005, \aap, 438, 875 

\end{thebibliography}
\end{document}